\documentclass[preprintnumbers,amsmath,amssymb,prl,superscriptaddress]{revtex4}

\usepackage{graphicx}  

\begin{document}

\title{Super KEKB ~/~ Belle II Project}
\author{B.~Golob}\affiliation{Faculty of Mathematics and Physics, University of Ljubljana,
  Ljubljana, Slovenia}\affiliation{J. Stefan Institute, Ljubljana, Slovenia} 

\collaboration{The Belle II Collaboration}



\begin{abstract}
We present the status of the KEKB accelerator and the Belle
detector upgrade, along with several examples of physics measurements 
to be performed with Belle II at Super KEKB. 
\end{abstract}

\maketitle

\section{Introduction}
\label{sect01}
The $B$ factories - the Belle detector taking data at the KEKB collider at KEK
\cite{Belle,KEKB} and 
the BaBar detector \cite{BaBar} at the PEP II at SLAC - 
have in more than a decade of data taking 
outreached the initial expectations on the physics results. They 
proved the validity of the Cabibbo-Kobayashi-Maskawa model of the 
quark mixing and $CP$ violation ($CPV$). Perhaps even more importantly,
they pointed out few hints of discrepancies between the Standard 
Model (SM) predictions and the results of the measurements. Facing the
finalization of the data taking operations the
question thus arises about the future experiments in the field 
of heavy flavour physics, to experimentally verify the current hints of
possible new particles and processes often addressed as the New
Physics (NP). Part of the answer are the planned Super $B$
factories in Japan and Italy, that could perform a highly sensitive
searches 
for NP, complementary to the long expected ones at the Large Hadron Collider. 
The so called precision frontier represented by the two machines
requires the achieved luminosities of the $B$ factories to be raised
by ${\cal{O}}(10^2)$. 
In the present paper we summarize the plan and the
status of the Belle detector upgrade (Belle II) at the upgraded KEKB 
(Super KEKB) $e^+e^-$ collider. 

In the following section we first briefly discuss the necessary
upgrade of the KEKB accelerator. In sections~\ref{sect03-1} to
\ref{sect03-3} we summarize the upgrade of the vital parts of the Belle
detector - the vertexing, the particle identification system and
the electromagnetic calorimeter, respectively. The upgrade is illustrated
with examples of planned measurements that will greatly benefit from the improved
collider and detector performance. Finally we draw short conclusions 
in Sect.~\ref{sect04}. 


\section{From KEKB to Super KEKB}
\label{sect02}

\begin{figure}
\includegraphics[width=1.0\textwidth]{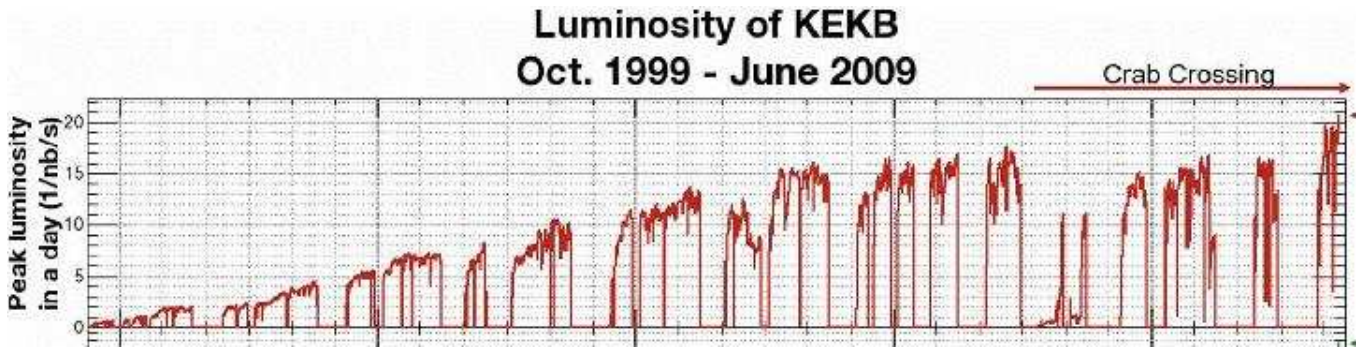}     
\caption{Daily peak luminosity of the KEKB collider.}
\label{fig01}
\end{figure}

The KEKB accelerator is an asymmetric $e^+e^-$ collider operating at
and near the center of mass energy of 10.58~GeV, corresponding to the
mass of the $\Upsilon(4S)$ resonance. The asymmetry of the beams
results in a Lorentz boost factor of $\beta\gamma=0.425$ which enables the time
dependent measurements in the system of $B$ mesons. The history of
the KEKB luminosity is presented in Fig.~\ref{fig01}. The highest
luminosity ever reached in the accelerator ($2.1\times
10^{34}$~cm$^{-2}$s$^{-1}$) 
is a result of the
crab cavities installed in 2007 \cite{crab_cav}. The continuous injection scheme and a
very stable operation made possible to collect data corresponding to
the integrated luminosity of more than 1~ab$^{-1}$.  

\begin{figure}
\begin{center}
\includegraphics[width=0.7\textwidth]{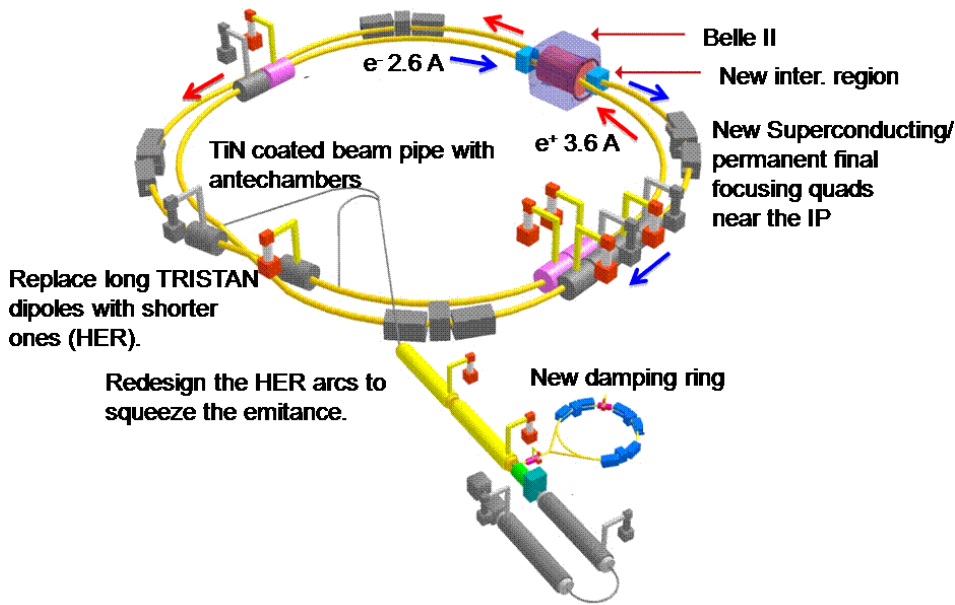}
\end{center}
\caption{The main parts of the KEKB upgrade.}
\label{fig02}
\end{figure}

The luminosity of the collider is governed by several factors. The
crucial ones for the upgrade of the KEKB are~\footnote{The subscripts
  $\pm$ denote the high energy electron and the low energy positron
  beam, HER and LER, respectively.} the beam currents ($I_\pm$), the
vertical beta function at the interaction point ($\beta_{y\pm}^\ast$) 
and the beam-beam
parameter $\xi_{y\pm}$. To start from the latter, the beam-beam parameter,
$\xi_{y\pm}=\sqrt{\beta_{y\pm}^\ast/\epsilon_y}$, will
remain almost unchanged at Super KEKB, $\xi_{y\pm}\sim 0.1$. The beta
function, however, will be extremely reduced: $\beta_{y\pm}^\ast=$~
5.9~mm/5.9~mm $\to$ 0.27~mm/0.41~mm.~\footnote{Due to the so called
  hourglass effect this requires also a reduction of the
  $\beta_{x\pm}^\ast$.} The emittance will be reduced accordingly to
match the current $\xi_{y\pm}$. Both beam currents will be also
increased by roughly a factor of two. In terms of the $e^+e^-$ bunches
the foreseen upgrade corresponds to the reduction of the current size
in direction perpendicular to the beam direction from $\sigma_x\sim
100~\mu$m, $\sigma_y\sim 2~\mu$m to $\sigma_x\sim
10~\mu$m, $\sigma_y\sim 60$~nm. To achieve the desired goal the main
tasks during the upgrade will be the installation of longer bending
radius in the LER, more arc cells in the HER, 
re-design of the interaction region with the new
final focusing quadrupoles closer to the interaction point, new beam
pipe and a new damping ring (see Fig.~\ref{fig02}). 
The outstanding problems are a rather
small dynamic aperture, larger Touschek background and consequently a
shorter lifetime of the beams, directly affecting the luminosity. To
cope with these, the upgrade includes an increased crossing angle of
the two beams (from 22~mrad to 83~mrad) and a slightly smaller
asymmetry of the beams (from 3.6~GeV/8~GeV to 4~GeV/7~GeV). 

\begin{figure}
\begin{center}
\includegraphics[width=0.7\textwidth]{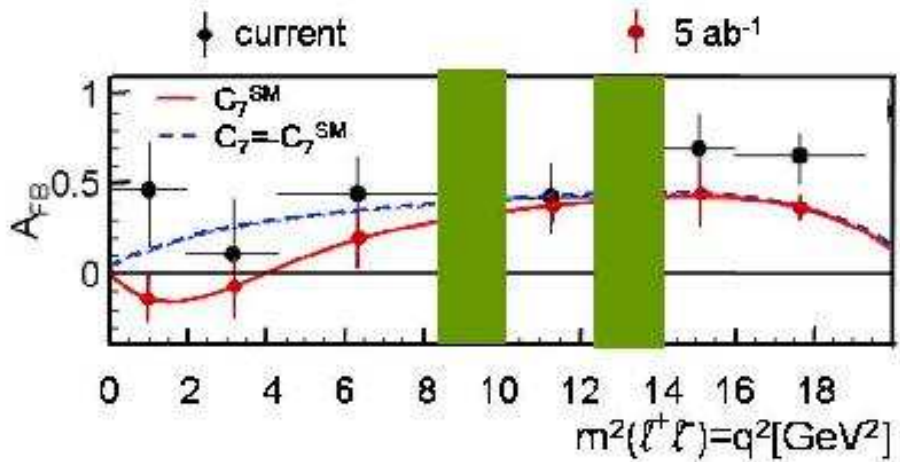}   
\end{center}
\caption{The measurement of the lepton forward-backward asymmetry in $B\to
K^\ast\ell^+\ell^-$ with 600~fb$^{-1}$ \cite{belle_afb} and
5~ab$^{-1}$ data (expected accuracies shown at the SM prediction). 
Shaded regions correspond to the charmonium veto $q^2$ intervals.}
\label{fig03}
\end{figure}

The luminosity of the Super KEKB will reach 
${\cal{L}}=8\times 10^{34}$~cm$^{-2}$s$^{-1}$. 
Assuming the startup of the machine in 2014, and a
rather conservative increase of the starting luminosity to the design
value, already in two years of data-taking the available data sample
will correspond to 5~ab$^{-1}$. Integrated luminosity of 50~ab$^{-1}$
is expected in 2020. To illustrate the precision that could be
achieved with such a large sample of $B$ meson decays we use the
measurement of the lepton forward-backward asymmetry $A_{FB}$ in $B\to
K^\ast\ell^+\ell^-$ decays. This observable (or even more so, the zero
crossing-point of the $A_{FB}(q^2)$, with $q^2\equiv m^2(\ell\ell)$)
is not very sensitive to the theoretical uncertainties arising from
the unknown form factors \cite{burdman_afb}. In
Fig.~\ref{fig03} the current Belle measurement \cite{belle_afb} is compared to the
expected sensitivity at Belle II with $\int{{\cal{L}}dt}=5$~ab$^{-1}$. 
It can be seen that such a
measurement will make possible a distinction among various models, for
example the SM and the Supergravity models with the reversed sign of
the $C_7$ Wilson coefficient.~\footnote{Note that this specific
  measurement can also be performed with a high precision at the
  LHCb. In the following we give examples of measurements that are
  completely complementary to NP searches at the LHC.}

\section{From Belle to Belle II}
\label{sect03}

\begin{figure}
\begin{center}
\includegraphics[width=0.5\textwidth]{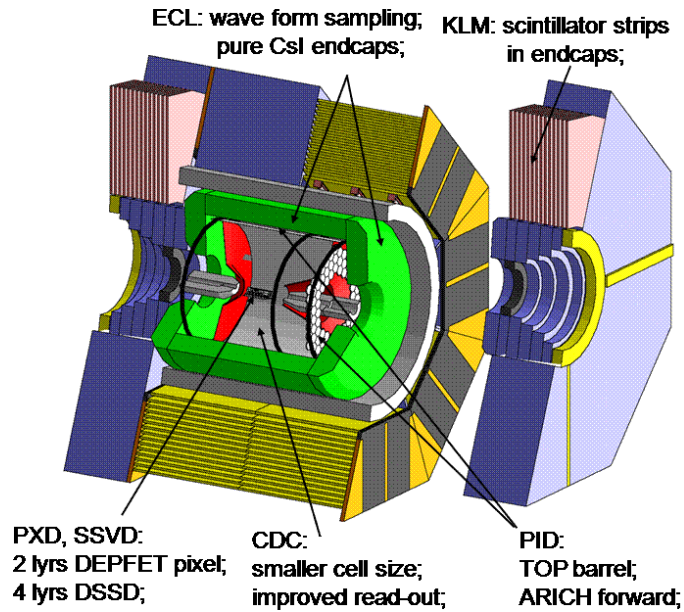} 
\end{center}
\caption{The main ingredients of the Belle detector upgrade.}
\label{fig04}
\end{figure}

A rough overview of the Belle detector upgrade is sketched in
Fig.~\ref{fig04}. In the environment of the beams with luminosity of 
${\cal{O}}(10^{35})$~cm$^{-2}$s$^{-1}$ the detector will have to cope
with an increased background (10-20 times compared to the present), 
which will be the cause of an increased occupancy and radiation
damage. 
The first level trigger rate is expected to increase from the current 0.5~kHz to
around 20~kHz. For several detector components we nevertheless foresee
an improved performance and a better overall hermiticity of the
detector after the upgrade. The task of vertexing will rely on two layers
of DEPFET pixel detectors (PXD) and four layers of double sided silicon
detectors (SSVD). The main tracking device, the Central Drift Chamber (CDC), will
have a smaller cell size and an improved read-out system. The particle
identification will be performed mainly by the Time-of-Propagation
counter (TOP) in the barrel and the RICH detector with aerogel radiator 
(ARICH) in the forward
part. For the electromagnetic calorimeter (ECL) the electronics enabling a
wave form sampling will be introduced, and some of the current CsI
crystals doped with Tl are going to be replaced by the pure CsI. The
detector of muons and $K_L$'s (KLM) will be upgraded with
scintillator strips in the endcaps. 

\subsection{Vertexing}
\label{sect03-1}

\begin{figure}
\includegraphics[width=0.5\textwidth]{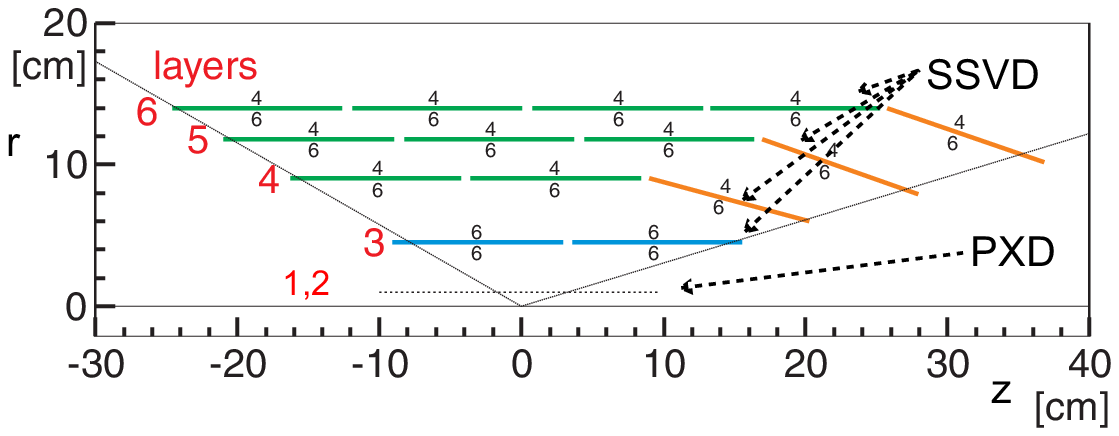} 
\includegraphics[width=0.4\textwidth]{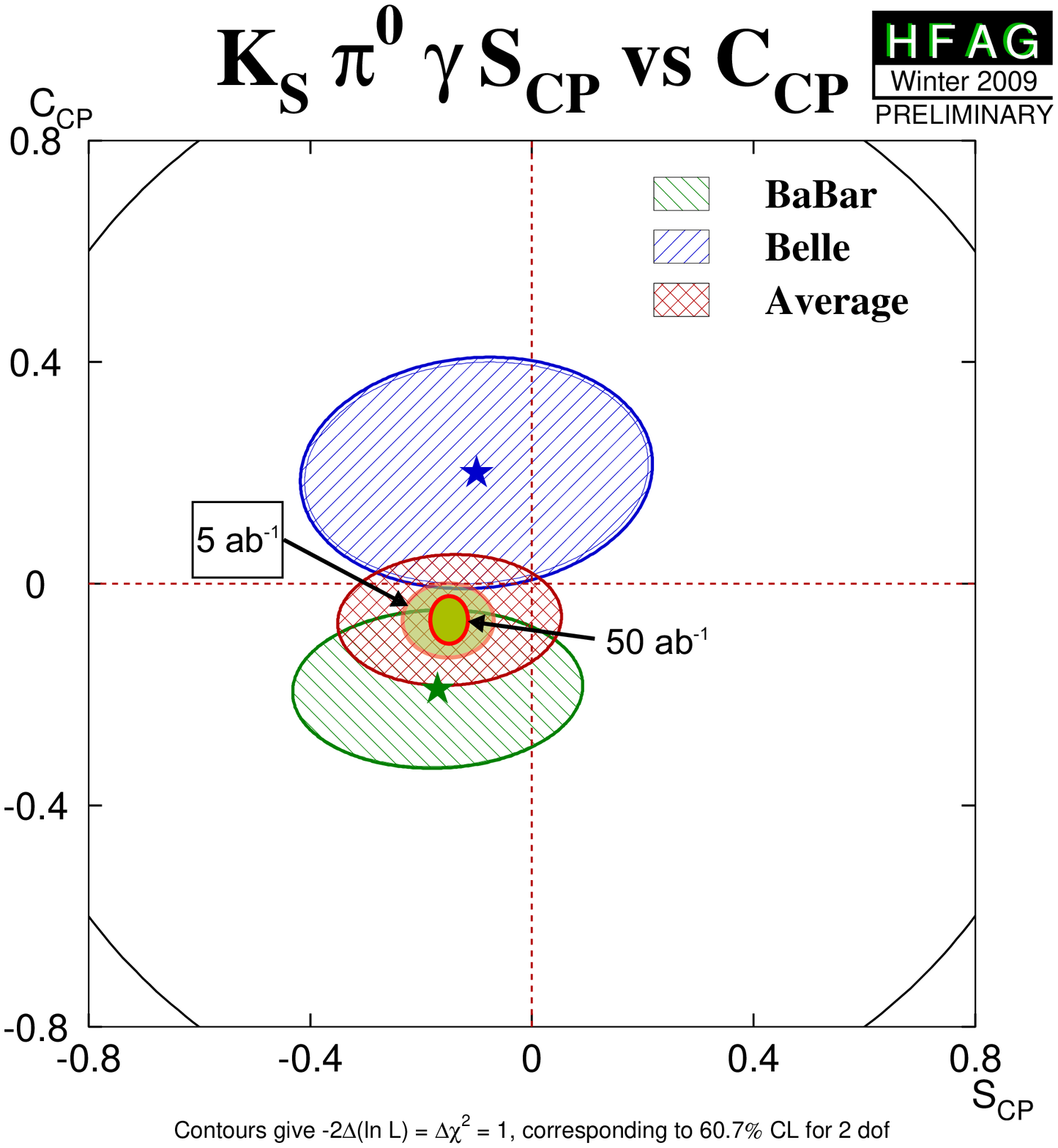}
\caption{Left: A schematic view of the upgraded semiconductor
  detector. Right: Comparison of the current \cite{hfag} and expected precision on
  direct and indirect $CPV$ in $B^0\to K_S\pi^0\gamma$.}
\label{fig05}
\end{figure}

A schematic view of the future semiconductor detector of Belle II is
shown in Fig.~\ref{fig05}(left) and is composed of two layers of pixel
detectors \cite{pixel} followed by four layers of double sided silicon
strip detectors. The improvement compared to the current
detector is twofold: a better spatial resolution of the vertex
determination (for around 25\% in the case of $B\to J/\psi K_S$ vertex), 
and an improved reconstruction efficiency of $K_S\to\pi^+\pi^-$ decays
with pion signals in the detector, due to the increased radii of the
layers. The latter is important for the time
dependent measurements of various decay modes with $K_S$'s in the
final state (increase of efficiency by around 30\%). Since the dependence on the
radius 
of the layers is opposite
for the two mentioned improvements, a careful optimization of the
design was performed. 

To illustrate the expected performance, a search for possible
right-handed currents in $B^0\to K_S\pi^0\gamma$ is used as a benchmark
mode. In these decays only the $K_S$ direction is used together with
the interaction point constraint to determine the $B$ meson decay
vertex \cite{belle_kastgam}. While the indirect $CPV$ is
heavily suppressed in the SM due to the helicity structure of the
Hamiltonian, it can be largely increased is some NP models
\cite{th_kastgam}. 
Figure~\ref{fig05}(right)
shows a comparison of the current values of the direct and indirect
$CPV$ parameters in this mode \cite{hfag} with the approximate expected precision
including the statistical and systematic uncertainties. The expected
accuracies on $S_{CP}$ with 5~ab$^{-1}$ and 50~ab$^{-1}$ are 0.09 and
0.03, respectively.   
While in the Left-Right Symmetric
Models $S_{CP}$ can be as high as 0.5, the sensitivity with 50~ab$^{-1}$ of
data is smaller than the SM predictions. 

\subsection{Particle identification}
\label{sect03-2}

Particle identification at Belle II will rely on the
TOP \cite{top} counter in the barrel part, and ARICH
detector in the forward \cite{arich}. The TOP detector will consist of 
quartz bars with mirrors on one side and microchannel plate
photomultipliers on the other. For high momentum (3 GeV/$c$) kaons we expect around
10\% better identification efficiency (90\%-95\%) in the barrel at a similar
misidentification probability as for the current 
detector (~5\%). 

\begin{figure}
\includegraphics[width=1.0\textwidth]{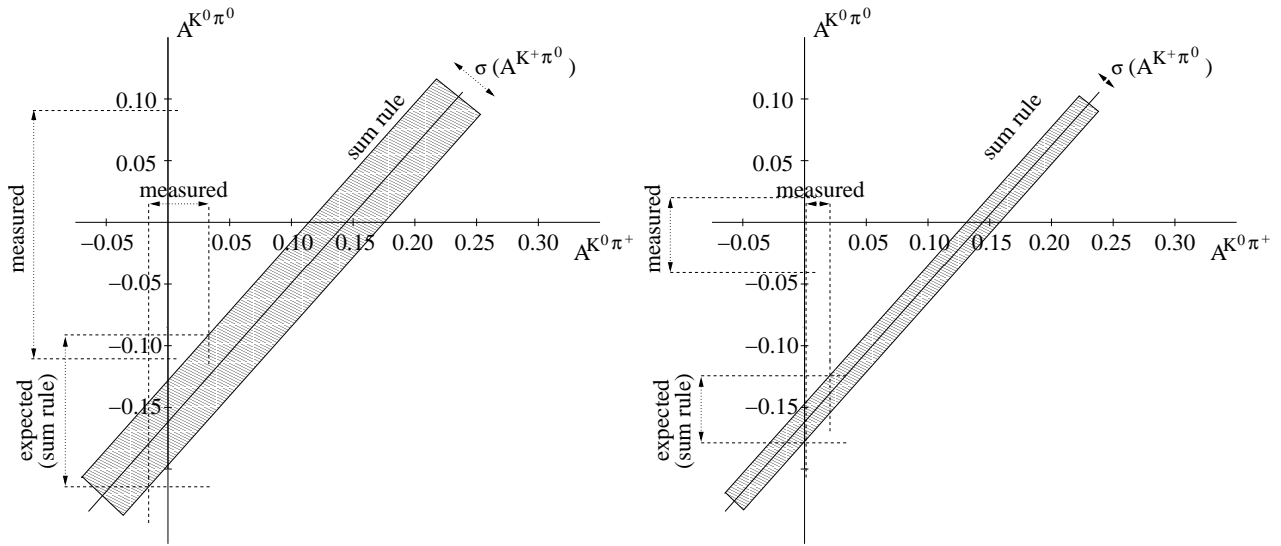}
\caption{Left: Direct measurements of $A_{K^0\pi^0}$ and
  $A_{K^0\pi^+}$ \cite{hfag} compared to the sum rule \cite{sum_rule}
  prediction. The width of the band is determined by the experimental uncertainties
  on other quantities entering the sum rule. Right: The same comparison
  assuming uncertainties expected with 50~ab$^{-1}$ of data.}
\label{fig06}
\end{figure}

Particle identification is of course crucial in several
measurements, for example the measurements related to the so called
direct $CPV$ puzzle, which arises from the observed difference between
the direct
$CPV$ asymmetry in $B^0\to K^+\pi^-$ and $B^+\to
K^+\pi^0$ decays~\cite{belle_akpi}. While in the explicit calculations of the 
asymmetries $A_{K\pi}$ several model uncertainties are present, a model
independent sum rule was proposed \cite{sum_rule} to test the
consistency of the SM. It relates the asymmetries and the branching
fractions of several decay modes:
$A_{K^+\pi^-}+A_{K^0\pi^+}[{\cal{B}}_{K^0\pi^+}/{\cal{B}}_{K^+\pi^-}][\tau_{B^0}/\tau_{B^+}] 
  = A_{K^0\pi^0}[2{\cal{B}}_{K^0\pi^0}/{\cal{B}}_{K^+\pi^-}]+ 
A_{K^+\pi^0}[{2\cal{B}}_{K^+\pi^0}/{\cal{B}}_{K^+\pi^-}][\tau_{B^0}/\tau_{B^+}]$. 
Figure \ref{fig06}(left) shows the current status of the measurements
using the world average values of the measured observables
\cite{hfag}, where $A_{K^0\pi^0}$ is expressed as a function of the
$A_{K^0\pi^+}$ 
using the
sum rule. The predictions of the sum rule are in agreement with the
direct measurements. The main systematic uncertainty on the 
$A_{K^0\pi^0}$, which is the least precisely known asymmetry in the
$K\pi$ system, can be reduced wit a larger data sample, and hence the
expected accuracy with 50~ab$^{-1}$ is 0.03. 
Assuming the current central values a 
discrepancy between the measurements and the sum rule
prediction at $\int{\cal{L}}dt=50$~ab$^{-1}$ would be significant (Fig.~\ref{fig06}(right)).   


\subsection{Electromagnetic calorimeter}
\label{sect03-3}

In the electromagnetic calorimeter (ECL) upgrade the replacement of the
current electronics is foreseen. The new one will enable amplitude-time
measurements for the signals in the ECL and will thus help to suppress
the background from clusters caused mainly by the off-time 
beam background (we expect the reduction of this background by a factor of 7). 
Beside this, a partial replacement of the Tl doped CsI
crystals with the radiation hard pure CsI is being under the consideration. 
Due to the increased rate of backgrounds at Belle II the expected
photon detection efficiency of the ECL is around 5\%-10\% lower,
while keeping the background at the current level. 

\begin{figure}
\begin{center}
\includegraphics[width=0.4\textwidth]{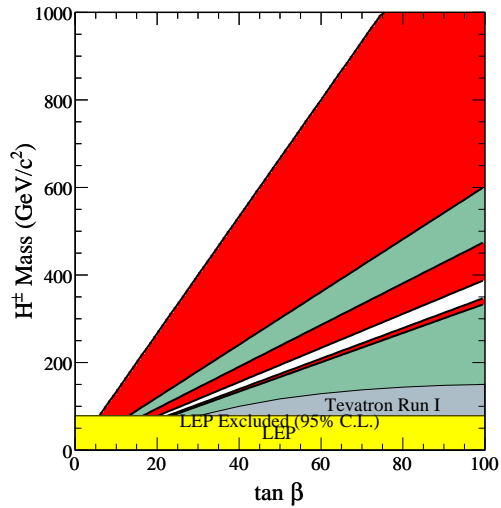}     
\end{center}
\caption{Five standard deviations discovery region (red, dark shaded) for the charged
  Higgs boson in the $(m_{H^\pm},\tan\beta)$ plane, from the
  measurement of ${\cal{B}}(B^+\to \tau^+\nu)$ with
  50~ab$^{-1}$ \cite{belle2_phys}. 
Other shaded regions show the current 95\% C.L. exclusion region.}
\label{fig07}
\end{figure}

The importance of the ECL performance can be best illustrated by the measurement of the 
${\cal{B}}(B^+\to \tau^+\nu)$ \cite{belle_taunu}. 
The method consists of full
or partial reconstruction of the tagging $B$ meson, identification of
hadrons or charged leptons from the $\tau$ decay, and examination of
the distribution of the remaining
measured energy in the event. For the signal, where the undetected
particles are neutrinos, this distribution of energy measured in the ECL
peaks at zero. The leptonic $B$ meson decays are interesting since,
for example, in the Type II Two Higgs Doublet Models, the SM branching
fraction receives a contribution from the charged Higgs boson
exchange, expressed as a multiplicative factor: ${\cal{B}}(B^+\to
\tau^+\nu)={\cal{B}}^{SM}[1-(m_B^2/m^2_{H^\pm})\tan^2\beta]^2$.  
The expected sensitivity on
${\cal{B}}(B^+\to\tau^+\nu)/{\cal{B}}^{SM}$ with 50~ab$^{-1}$ of data
is below 0.1. 
With an increased statistical power of the data, and
assuming the existing ECL performance, one obtains the five standard
deviations discovery region for the charged Higgs boson as shown in
Fig.~\ref{fig07} \cite{belle2_phys}. It can be seen that a large area of the
$(m_{H^\pm},\tan\beta)$ plane (compared to the current exclusion
regions) can
be covered by this search, especially at larger values of the mass and
$\tan\beta$ \footnote{In the estimation of the expected sensitivity we
  also assumed an improvement in the $|V_{ub}|$ and $f_B$ values
  precision to $\pm$3\% each.}. Such a measurement is to some extent complementary to
the measurements of the $b\to s\gamma$ transition branching fraction which
constrain the mass of the charged Higgs boson almost independently of
the $\tan\beta$ value. 

\section{Summary}
\label{sect04}

In summary, we presented a short overview of the KEKB accelerator and
Belle detector upgrade. While technologically most challenging, the
preparation of the Super $B$ factory at KEK is well on the way. The key
features of the upgrade are illustrated by several measurements that
will be possible at Belle II and will represent a highly sensitive
search for NP effects, complementary to searches at the LHC. A
comprehensive program of physics measurements 
can be found in \cite{belle2_phys}. 

\begin{figure}
\begin{center}
\includegraphics[width=0.45\textwidth]{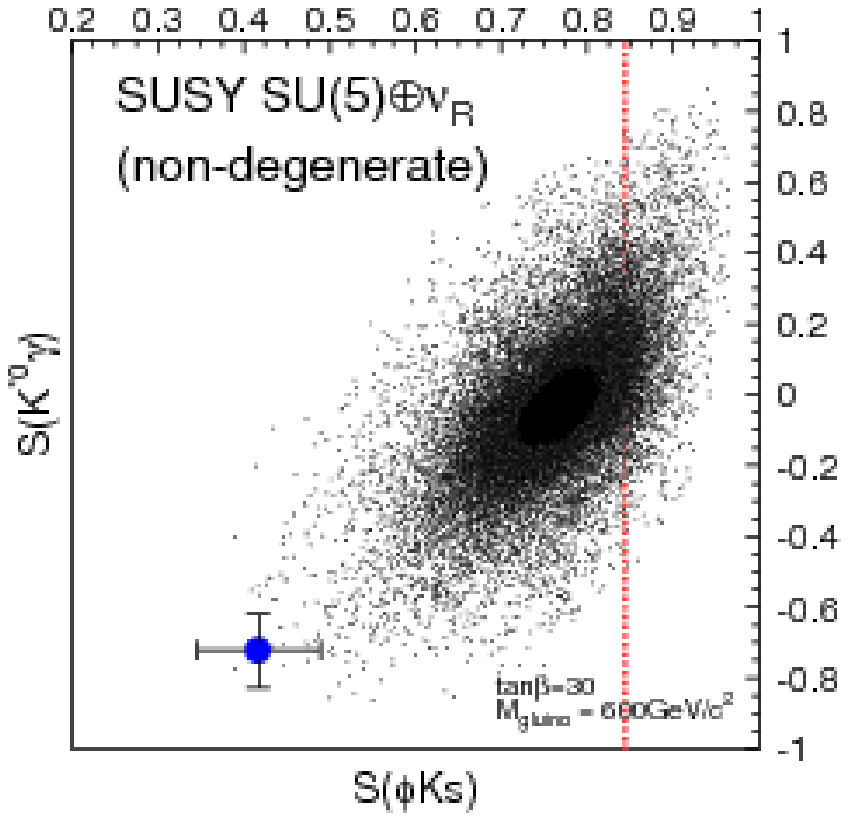}
\includegraphics[width=0.45\textwidth]{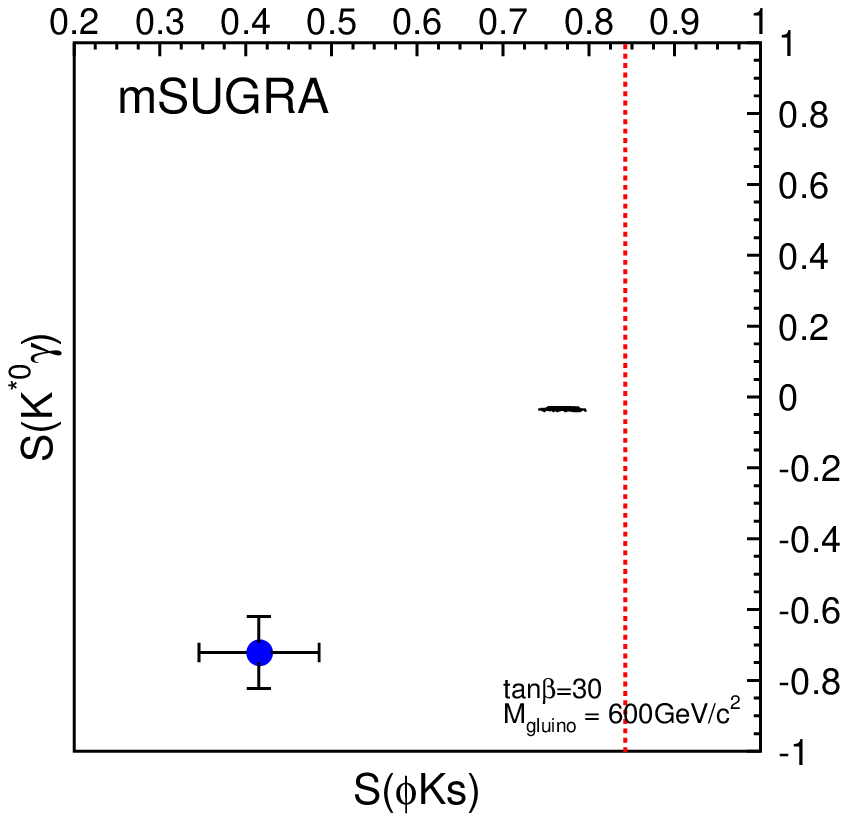}
\end{center}
\caption{Comparison of the correlation between the indirect $CPV$ parameter in 
$B^0\to K^{\ast 0}(\to K_S\pi^0)\gamma$ and $B^0\to\phi K_S$
  decays~\cite{belle2_phys} for two models: the Minimal super
gravity model (mSUGRA) and Supersymmetric grand unification theory
with right-handed neutrinos (SUSY SU(5)). 
The points with error bars denote the expected sensitivity
  at Belle II with 5~ab$^{-1}$ of data. }
\label{fig08}
\end{figure}

The value of the Super $B$ factory lies not only in a highly sensitive
search of NP in individual processes, but to a large extent in the
possibility of performing measurements of various observables which
through their correlations can help identifying the nature of NP. As
an example, Fig.~\ref{fig08}~\cite{belle2_phys} shows correlations between the indirect
$CPV$ in $B^0\to K^{\ast 0}(\to K_S\pi^0)\gamma$ (see
Sect.~\ref{sect03-1}) and in the $B^0\to\phi K_S$ with the underlying
penguin quark process $b\to s\bar{s}s$ (a naive SM prediction is
$S_{\phi K_S}=S_{J/\psi K_S}=\sin{2\phi_1}$). In the Minimal super
gravity model (mSUGRA) and Supersymmetric grand unification theory
with right-handed neutrinos (SUSY SU(5)) the correlations between the two
observables exhibit a different pattern, while the mass spectra of
the particles predicted in the two models are similar. If hints of new
particles consistent with these predictions arise from the LHC, at Belle II one can
distinguish the two models with already $\int{\cal{L}}dt=5$~ab$^{-1}$ of data. 

The preparation of the Belle II
detector at Super KEKB is proceeding according to the plan, with the aim of
starting the data taking in 2014. 

\end{document}